\documentclass[12pt]{article}
\usepackage{graphicx}
\usepackage{amsmath}
\usepackage{amsfonts}
\usepackage{amssymb}
\newtheorem{theorem}{Theorem}[section]

\newtheorem{corollary}[theorem]{Corollary}
\newtheorem{remark}[theorem]{Remark}
\newenvironment{proof}[1][Proof]{\textsc{#1.} }{\ \rule{0.5em}{0.5em}}
\numberwithin{equation}{section}

\begin{document}

\title{Motion of Isolated bodies}

\author{\and Yvonne Choquet-Bruhat\\ 
Acad\'emie des Sciences \\
23 Quai de Conti\\
75270 Paris Cedex 06, France\\
and\\ 
Helmut Friedrich\\ 
Max-Planck-Institut f\"ur Gravitationsphysik\\
Am M\"uhlenberg 1\\
14476 Golm, Germany}

\maketitle

\begin{abstract}
It is shown that sufficiently smooth initial data for the Einstein-dust or the Einstein-Maxwell-dust equations with non-negative density of compact support develop into solutions representing isolated bodies in the sense that the matter field has spatially compact support and is embedded in an exterior vacuum solution.  

\end{abstract}


\section{Introduction}

While a detailed understanding of the motion of massive isolated bodies in
general relativity is, for obvious physical reasons, of great interest, only a
few special situations have been successfully analysed so far.

The initial value problem for the Einstein equations with suitable matter
fields, such as a perfect fluid model with an appropriate equation of state, is
well understood in domains where the total energy density is positive. Domains
with vacuum-matter interfaces, which define the boundaries of the bodies,
still pose technical problems. While quite general initial data for the
Einstein-Euler system have been constructed by Dain and Nagy (\cite{dain:nagy}%
), which describe compact fluid balls embedded in an exterior vacuum field,
the evolution problem for these data has not been solved yet in the same generality.

If the density has a positive one-sided limit at the matter boundaries, where
the pressure is required to vanish, the right hand side of Einstein's field
equations will have discontinuities along the boundaries. Under the assumption
of spherical symmetry, which is very special from the PDE point of view, the
evolution problem has successfully been analysed in the presence of
discontinuities by Ehlers and Kind ( \cite{kind93}). Nothing is known for
situations with lower degrees of symmetry.

Situations where the density goes to zero at the boundaries with a certain
smoothness have been discussed by Rendall (\cite{rendall:I}). While no
symmetries are assumed, his method requires the use of particular equations of
state. These allow him to construct smooth solutions to the Einstein-Euler
system with fluid bodies of spatially compact support, but, as pointed out by
the author himself, the method he uses has undesirable properties such as a
lack of uniqueness and the impossibility to discuss perturbations of static
spherically symmetric solutions to the Einstein-Euler system.

The present article deals with a further case in which the density goes to
zero at the boundaries of the bodies and where no space-time symmetries are
required. It assumes pressure free matter (also referred to as `dust') or
charged pressure free matter as a matter model.

Two methods are known to write the Einstein-dust equations in hyperbolic
form. In \cite{friedrich98} have been employed the frame formalism and the
Bianchi equations for the conformal Weyl tensor to combine the hyperbolicity
of the equations with a Lagrangian representation of the fluid flow. This is
clearly of interest if one wants to control the location of the boundaries
under general assumptions on the fluid model. These equations appear to be
useful, however, only in four space-time dimensions.

The representation of \cite{choquet58} starts from the Einstein equations in
wave (`harmonic' in the older terminology) coordinates. To render the complete
system hyperbolic, a further derivative is applied to the equations for the
metric coefficients so that one ends up again with a system of third order in
the metric. The equations so obtained are applicable in all space-time
dimension. Since the location of the boundary is not an essential problem in
the situation considered here these equations will be used in the following.

In the existence proof we find it technically convenient to use a
`non-physical' extension of the matter flow vector field into the vacuum part
of the solution. It is then shown that the solutions are in fact independent
of these extensions and the latter can be ignored. Our main result concerning
the motion of general relativistic bodies consisting of pressure free matter
is then stated in the Theorem \ref{physexun}. The analogues result concerning
charged pressure free matter is given in Theorem \ref{phycharsexun}.

\section{Solutions to the Einstein-dust equations}

\subsection{The equations}

The stress energy tensor $T$ of pressure-free matter is
\begin{equation}
T_{\alpha\beta}=ru_{\alpha}u_{\beta}%
\end{equation}
where $r$ is the scalar matter density, and $u$ is the (future directed) flow
vector field which satisfies
\begin{equation}
u^{\alpha}u_{\alpha}=-1,\text{ hence \ \ }u^{\alpha}\nabla_{\beta}u_{\alpha
}=0,
\end{equation}
with $\nabla$ the covariant differential in the spacetime metric $g.$

The tensor $T$ satisfies the conservation law
\begin{equation}
\label{conslaw}
\nabla_{\alpha}T^{\alpha\beta}=0.
\end{equation}
Its projection orthogonal to $u$, which reads $r\,u^{\alpha}\nabla_{\alpha
}u_{\beta}=0$, implies near points where $r \neq0$ the geodesic equation
\begin{equation}
\label{ugeod}
u^{\alpha}\nabla_{\alpha}u_{\beta}=0.
\end{equation}
Its projection into the direction of $u$ gives the continuity equation
\begin{equation}
\label{contequ}
\nabla_{\alpha}(ru^{\alpha}) = u^{\alpha}\partial_{\alpha}r+r\nabla_{\alpha
}u^{\alpha}=0.
\end{equation}
The Einstein equations with pressure-free matter source read in $n+1$
dimensional spacetime
\begin{equation}
\label{einstdust}
R_{\alpha\beta}=r(u_{\alpha}u_{\beta}+\frac{1}{n-1}g_{\alpha\beta}).
\end{equation}
We shall regard equations (\ref{ugeod}), (\ref{contequ}), (\ref{einstdust}) as
our basic system of differential equations.

\subsection{An existence and uniqueness theorem.}

The geometric initial data for the spacetime metric $g$ on an initial manifold
$M$ are a Riemannian metric $\bar{g}$ and a symmetric 2-tensor $\bar{k}$. The
initial data for a dust source are a scalar function $\bar{r}$ on $M$ and a
tangent vector field $\bar{v}$ to $M.$ A solution $(V,g,r,u)$ of the coupled
Einstein-dust equations is an Einsteinian development of the initial data
set $(M,\bar{g},\bar{k},\bar{r},\bar{v})$ if $M$ can be diffeomorphically
identified with an embedded submanifold of $(V,g)$ so that $\bar{g}$ and
$\bar{k}$ are respectively the induced metric and second fundamental form on
$M$, while $\bar{r}$ is the function induced by $r$ on $M $ and $\bar{v}$ is
the value on $M$ of the dust velocity with respect to the proper frame (and
the proper time) of an observer with timelike vector
orthogonal\footnote{{\footnotesize Two such timelike vectors corresponding to
two developments isometric under a diffeomorphism f are mapped onto each other
by f.}} to $M$ in $(V,g).$ In local coordinates such that the values on $M$ of
the shift and the lapse of the development are respectively $\bar{\beta}=0 $
and $\bar{N}=1,$ it holds that $\bar{v}^{i}=(\bar{u}^{0})^{-1}\bar{u}^{i},$
where $\bar{u}^{\alpha}$ are the components of $u$ in the considered
coordinate system at points of $M.$

If $M$ is compact, we denote by $H_{s}$ a usual Sobolev (Hilbert) space of
functions or tensor fields on M which are square integrable together with
their derivatives of order up to s in a given smooth Riemannian metric on the
manifold M. The notation $M_{s}$ stands for continuous and bounded Riemannian
metrics with derivatives in $H_{s-1}$. On non compact manifolds suitable
variants of these spaces can be used. Since the equations we consider are
hyperbolic, the evolution problem can be localized. The discussion of the
basic problem, namely the analysis of the evolution in domains containing
boundaries of the bodies, will thus be essentially the same for compact or
non-compact initial manifolds\footnote{{\footnotesize One can replace, for
instance, the spaces H}$_{s}${\footnotesize \ \ by spaces H}$_{s}^{u.loc}%
${\footnotesize \ \ of tensors with belong to H}$_{s}${\footnotesize \ \ in
the open sets \ of a locally finite covering of the manifold M, with uniformly
bounded H}$_{s}${\footnotesize \ norms.}}.

A tensor field of degree $p$ on a manifold $M\times R,$ with $R$ parametrized by $t$,
can be decomposed, by projections on $R$ or $M,$ into a set ot $t-$dependent
tensor fields, of degree $0,1, \ldots ,p$ on $M.$ For a $t$-dependent a tensor
field on $M,$ $t\in\lbrack0,T[,$ the space $E_{s}(T)$ is defined as follows:
\[
E_{s}(T)=C^{0}([0,T[,H_{s})\cap C^{1}([0,T[,H_{s-1})\cap C^{2}([0,T[,H_{s-2}).
\]
We say that a tensor on $M\times\lbrack0,T[$ is in $E_{s}(T)$ if each of its
projections is in $E_{s}(T).$ We denote by $L_{s}(T),$ respectively by
$U_{s}(T),$ the space of Lorentzian metrics, respectively the space of unit
vectors in the metric $g,$ which belong to $E_{s}(T).$

We will prove the following local existence and geometrical global uniqueness theorem.

\begin{theorem}
\label{exun}
The Einstein equations in wave gauge with source a pressure free matter, form
a hyperbolic Leray system for $g,$ $u$ and $r,$ which is causal as long as $g$
is Lorentzian and $\ u$ is timelike. There is an interval $[0,T[\,\subset
\mathbb{R}$ such that the Cauchy problem for these equations with data on the
manifold $M$ with $\bar{g}\in M_{s}$, $\bar{k}\in H_{s-1}$, $\bar{v}\in
H_{s-1}$, $\bar{r}\in H_{s-2}$, where $s>\frac{n}{2}+2$, $|\bar{v}|_{\bar{g}%
}<1$, has one and only one solution $g\in L_{s}(T)$, $u\in U_{s-1}(T)$, $r\in
E_{s-2}(T)$.
\end{theorem}

\begin{corollary}
\label{geomunique}
If the initial data satisfy the Einstein constraints, the solution obtained in
wave gauge satisfy the original Einstein-dust system. The solution is
globally hyperbolic. There is a unique solution, up to isometry, in the class
of maximal globally hyperbolic spacetimes if $s>\frac{n}{2}+3.$
\end{corollary}

\begin{proof}
The Einstein equations with dust source read in wave coordinates:
\begin{equation}
\label{redeinst}
R_{\alpha\beta}^{(h)}\equiv-{\frac{1}{2}}g^{\lambda\mu}\partial_{\lambda\mu
}^{2}g_{\alpha\beta}+H_{\alpha\beta}(g,\partial g)=\rho_{\alpha\beta}\equiv
r(u_{\alpha}u_{\beta}+\frac{1}{n-1}g_{\alpha\beta}).
\end{equation}
They are to be coupled with equations (\ref{ugeod}), (\ref{contequ}).

We think of $M$ as being embedded into the solution manifold and the
coordinate $x^{0}$ to be chosen such that $M=\{x^{0}=0\}$. The solution
manifold then takes close to $M$ the form $\mathbb{R}_{0}^{+}\times M$, with
$x^{0}$ inducing a coordinate on the first factor, and $M\simeq\{0\}\times M
$. The initial values for the unknowns $g_{\alpha\beta}$ and $u^{\alpha} $ are
deduced from the geometric initial data set by assuming the contracted
Christoffel symbols $\Gamma^{\alpha}=g^{\beta\gamma}\,\Gamma_{\beta}%
\,^{\alpha}\,_{\gamma}$ to vanish on $M$ and the wave coordinates to be chosen
such that the values on $M$ of the lapse and the shift of the metric are
respectively $\bar{N}=1$ and $\bar{\beta}=0$.

We take a derivative of equations (\ref{redeinst}) in the direction of $u$ and
use equations (\ref{ugeod}), (\ref{contequ}) to obtain equations which are of
third order in $g$ but still do not contain derivatives of $r$:
\begin{equation}
\label{derredeinst}
u^{\gamma}\nabla_{\gamma}R_{\alpha\beta}^{(h)}=-r\,(u_{\alpha}u_{\beta}%
+\frac{1}{n-1}g_{\alpha\beta})\nabla_{\gamma}u^{\gamma}.
\end{equation}
The initial values for the second derivatives of $g$ are determined on $M$ so
that equations (\ref{redeinst}) will hold on $M$. This will imply that the
derivatives of the contracted Christoffel symbols $\Gamma^{\alpha}$ calculated
from the derivatives of $g$ will also vanish on $M$.

To the equations (referred to below by their equation numbers) we assign the
Leray-Volevic indices \cite{leray53}
\[
m(\ref{derredeinst})=0,\text{ \ \ }m(\ref{ugeod})=1,\text{ \ \ }%
m(\ref{contequ})=0,
\]
and to the unknowns $g$, $u$, $r$ the indices
\[
\ell(g)=3,\text{ \ }\ell(u)=2,\text{ \ }\ell(r)=1.
\]
The matrix of principal parts of the various orders $\ell-m$ is then diagonal
and given by
\[
\left(
\begin{array}
[c]{ccc}%
-{\frac{1}{2}}g^{\lambda\mu}u^{\gamma}\partial_{\gamma\lambda\mu}^{3}%
g_{\alpha\beta} & 0 & 0\\
0 & u^{\alpha}\partial_{\alpha}u^{\beta} & 0\\
0 & 0 & u^{\alpha}\partial_{\alpha}r
\end{array}
\right)  .
\]
The system is quasidiagonal with characteristic cotangent cone
\[
g^{\lambda\mu}\xi_{\lambda}\xi_{\mu}u^{\alpha}\xi_{\alpha}=0,
\]
the union of the light cone and of a spacelike hyperplane exterior to it, if
$u$ is a timelike vector. The system is thus hyperbolic and causal. The
coefficients of the principal terms are in $C^{1}$ under the given hypothesis.
The Leray-Dionne theory \cite{dionne}, \cite{leray53} gives, after some
work\footnote{The lowering of the regularity of the data required by the
general Leray theory of hyperbolic systems, which was worked out by Leray's
student Dionne, was written up only in the case of one equation, though of
arbitrary order.}, the existence of a number $T>0$ such that the considered
Cauchy problem has a unique solution $g\in L_{s}(T),$ $u\in U_{s-1}(T),$ $r\in
E_{s-2}(T).$

Since equation (\ref{derredeinst}) can then be rewritten as an ODE for
$R_{\alpha\beta}^{(h)}-\rho_{\alpha\beta}$ along the integral curves of $u$
and we have (\ref{redeinst}) arranged to be satisfied on $M$, it follows that
(\ref{redeinst}) is satisfied on the solution manifold.

The proof of the corollary follows the same lines as in the vacuum case.
\end{proof}

\subsection{The motion of isolated bodies.}

We have just proven that the Cauchy problem with data $\bar{g},\bar{k},\bar
{v},\bar{r}$ as in Theorem $\ref{exun}$ is geometrically well posed and
provides a unique solution $g$, $u$, $r$ on $V=[0,T[\times M$. There remains,
however, an open problem.

Denote by $\omega$ the maximal open subset of $M$ (assumed to be non-empty) on
which $\bar{r} > 0$. Suppose that its closure satisfies $\bar{\omega} =
supp(\bar{r})$ and, that $\bar{r}$ vanishes on an open set so that $M
\setminus\bar{\omega} \neq\emptyset$. We can think then of $\bar{\omega}$ as a
union of disjoint compact subsets of $M$ which represent the space occupied by
material bodies whose density tends continuously to zero at the boundary,
since $\bar{r}\in H_{2}$. While $\bar{v}$ represents the physically well
defined `matter flow velocity' at points of $\omega$, it has no physical
meaning at points where $\bar{r}$ vanishes (except, perhaps, on the boundary of
$\omega$ as limit of physical velocities). In fact, the $H_{s-1}$ extension of
$\bar{v}$ from $\omega$ to all of $M$ has only be introduced as a convenient
device to get the existence result.

Equation (\ref{contequ}) implies that $r=0$ along geodesics which start with tangent 
vectors $\bar{u}(p)$ with $p \in M \setminus \omega$. Along the geodesics with tangent 
vectors $\bar{u}(p)$ with $p \in \omega$ the function  $r$ will be positive and bounded 
as long as the convergence $- \nabla_{\alpha}u^{\alpha}$ remains bounded. If 
$- \nabla_{\alpha}u^{\alpha} \rightarrow \infty$, which indicates the development 
of a caustic of the geodesic vector field, we can expect $r \rightarrow \infty$ and the 
evolution comes to an end. For geometric reasons we cannot have 
$\nabla_{\alpha}u^{\alpha} \rightarrow \infty$, whence $r \rightarrow 0$,  in the future 
development.

The closure of the `geodesic tube' $\Omega$ of points in $V$ swept out by the geodesics
with origin in $\omega$ coincides with the support of $r$ in $V $. It
represents the history of the material bodies. In $V\setminus\Omega$ the
metric $g$ satisfies Einstein's vacuum field equations.

\begin{theorem}
\label{physexun} 
Suppose $\bar{r} \ge 0$. If $s>\frac{n}{2}+2$ the physical solution in wave gauge
obtained in Theorem \ref{exun}, which is given by the fields $g$ and $r$ on
$V$ and the field $u$ on $\Omega$, is uniquely determined in a neighbourhood
of $M$ by the data $\bar{g}$, $\bar{k}$, $\bar{r}$ on $M$ and the datum
$\bar{v}$ on ${\omega}$ with $\bar{\omega}=supp(\bar{r})$. It does not depend
on the extension of $\bar{v}$ to $M$.
\end{theorem}

\begin{remark}
The following discussion will also show that the life-time of the solution
considered in Theorem \ref{exun} might be increased by suitable redefinitions
of the field $u$ outside $\Omega$.
\end{remark}

\begin{proof}
We note that the smoothness result of the theorem (ensured by choosing $T$
small enough) excludes that any two of the geodesics corresponding to $u$
cross each other on $V$ (otherwise $u$ would not even correspond to a well
defined vector field). It follows in particular that none of the geodesics
which start at points of $M\setminus\omega$ enters the set $\Omega$. Because
$\Omega$ is generated by time-like geodesics starting at points of $\omega$,
it has empty intersection with the future $g$-domain of dependence
$D^{+}(M\setminus\bar{\omega})$ in $V$ of the set $M\setminus\bar{\omega}$. It
follows from general results about the Einstein equations that $g$ is
determined on $D^{+}(M\setminus\bar{\omega})$ uniquely by the restriction of
the data $\bar{g}$ and $\bar{k}$ to $M\setminus\bar{\omega}$.

Suppose, $\bar{v}_{\ast}$ is a datum on $M$ analogues to $\bar{v}$ so that
$\bar{v}_{\ast}=\bar{v}$ (whence $\bar{u}_{\ast}=\bar{u}$) on $\omega$ and
$\bar{g},\bar{k},\bar{v}_{\ast},\bar{r}$ satisfy the requirements of Theorem
\ref{exun} with $s>\frac{n}{2}+2$.
Then there exists a
number $T_{\ast}$, $0<T_{\ast}\leq T$, so that the tangent vectors of the
$g$-geodesics in $V$ with tangent vector $\bar{u}_{\ast}$ on $M$ define a
non-vanishing vector field $u_{\ast}$ on $V_{\ast}=[0,T_{\ast}[\times M\subset
V$. 

It follows that for points $p\in\omega$ the restrictions to $V_{\ast}$ of the 
$g$-geodesics with initial vectors $\bar{u}_{\ast}(p)$ resp. $\bar{u}(p)$ coincide 
and are contained in $V_{\ast} \cap \Omega$.
Equation (\ref{contequ}) with $u$ replaced by $u_{\ast}$ then determines a 
unique solution $r_{\ast}$ on $V_{\ast}$ which agrees with $\bar{r}$ on\textbf{\ }$M$.
The functions $r$ and  $r_{\ast}$ have their support in the closure of $V_{\ast} \cap \Omega$,
and we have in fact $r_{\ast}=r$ everywhere on $V_{\ast}$.

The expression on the right hand side of (\ref{derredeinst}) remains unchanged
if we replace $u$ and $r$ by $u_{\ast}$ and $r_{\ast}$. The fields $g$,
$u_{\ast}$, $r$ thus define on $V_{\ast}$ a solution to the Einstein-dust
equations. By the uniqueness statement of Theorem \ref{exun} it must coincide
with the solution $g_{\ast},u_{\ast},r_{\ast}$ determined by the data $\bar{g},
\bar{k},\bar{v_{\ast}},\bar{r}$ on $M$. Therefore $g=g_{\ast}$ and
$r_{\ast}=r$ on $V_{\ast},$ and, as seen previously $u=u_{\ast}$ on $V_{\ast
}\cap\Omega.$ The data $\bar{g},\bar{k},\bar{v},\bar{r}$ and $\bar{g},\bar
{k},\bar{v_{\ast}},\bar{r}$ determine the same physical solution on $V_{\ast}$.

The argument would be complete if we had $T_{\ast}=T$ for all possible
extensions of $v|_{\omega}$ to $M$. This can, however, not be expected. Since
geodesic flows are defined by equations of second order, they tend to develop
caustics. If this happens with the geodesics
generating $\Omega$ this might lead, as remarked above,  to a blow up of the density $r $ and to
physical phenomena like shell crossings which have been studied in the
literature (cf. \cite{yodzis:seifert:muellerzumhagen}).

Whether geodesics outside $\Omega$ show this behaviour or whether geodesics
starting on $M\setminus\bar{\omega}$ tend to approach $\Omega$ depends on the
metric $g$ as well as on the choice of extension of $v|_{\omega}$ to $M$, and
the latter will control to some extend the location of the set where this is
going to happen. By a more judicious choice of the extension or by subsequent
redefinitions of the extensions on suitable slices $\{t=const.\}$ this
phenomenon, which is of no physical or geometrical relevance but may restrict
the domain of existence in a similar way as a bad choice of gauge, can be avoided.
\end{proof}

The Einstein-dust constraints on $M$ do not depend on the extension of
$\bar{v}$ at points where $\bar{r}$ vanishes. They depend only of $\bar{g}$,
$\bar{k},$ $\bar{r}$ on $M$ \ and $\bar{v}$ on $\bar{\omega}$ if 
$supp(\bar{r}) = \bar{\omega}$. We can therefore state the following corollary.

\begin{corollary}
\label{physunique}
If the initial data $\bar{g}$, $\bar{k},$ $\bar{r}$ on 
$M$  and $\bar{v}$ on $\bar{\omega} = supp(\bar{r})$ 
satisfy the Einstein - dust
constraints, then the physically unique local solution obtained in wave gauge
in theorem \ref{exun} satisfies the full Einstein equations. It is locally
geometrically unique.
\end{corollary}

\begin{proof}
Since any extension to $M$ of the given initial data satisfies the constraints, 
Corollary \ref{geomunique} shows that the corresponding solution on $V$ obtained in wave
gauge satisfies the full Einstein-dust equations. The same holds a fortiori
for the restriction of this solution to the physically unique solution.

To prove local geometric uniqueness we consider a solution of the Einstein-dust 
system $g,r,u,$ in arbitrary coordinates, with Cauchy data $\bar{g}$,
$\bar{k},$ $\bar{r}$ on $M,$ and $\bar{v}$ on
$\bar{\omega} = supp(\bar{r})$. One can show that there exist in a neighbourhood of $M$ a
change to harmonic coordinates reducing to the identity on $M,$ hence
preserving $\bar{r}$ and $\bar{v}$, and such that $\bar{g},$ $\bar{k},$ are preserved.
\end{proof}

\section{Charged dust.}

\subsection{The equations.}

The stress energy tensor of charged pure matter (dust) is the sum of the
stress energy tensor of the matter and the Maxwell tensor of the
electromagnetic field $F:$%
\begin{equation}
\label{charseten}
T_{\alpha\beta}=r\,u_{\alpha}u_{\beta}+\tau_{\alpha\beta},
\end{equation}
with
\begin{equation}
\tau_{\alpha\beta}=F_{\alpha}{}^{\lambda}F_{\beta\lambda}-\frac{1}{4}%
g_{\alpha\beta}F^{\lambda\mu}F_{\lambda\mu}.
\end{equation}
The Einstein equations read
\begin{equation}
\label{chareinstdust}
R_{\alpha\beta}=\Phi_{\alpha\beta}+r(u_{\alpha}u_{\beta}+\frac{1}%
{n-1}g_{\alpha\beta})
\end{equation}
where we have set
\begin{equation}
\Phi_{\alpha\beta}:=F_{\alpha}{}^{\lambda}F_{\beta\lambda}+c_{n}g_{\alpha
\beta}F^{\lambda\mu}F_{\lambda\mu},\text{ \ \ }c_{n}:=(\frac{n-3}{n-1}%
-\frac{1}{4}).
\end{equation}
The Maxwell equations are, with $J$ the convection electric current of the
density of charge $q$
\begin{equation}
dF=0\text{ \ \ and \ \ }\nabla.F=J,\text{ \ i.e. \ \ }\nabla_{\alpha}%
F^{\alpha\beta}=J^{\beta}:=qu^{\beta};
\end{equation}
they imply the conservation of charge equation
\begin{equation}
\label{charcons}
\nabla_{\alpha}(qu^{\alpha})=0.
\end{equation}

For simplicity we suppose the space-time to be simply connected so that there
exists a 1-form $A$, the electromagnetic potential, with
\begin{equation}
F=dA.
\end{equation}
We take $A$ in the Lorentz gauge, i.e. so that
\begin{equation}
\delta A=0,\text{ \ i.e. \ }\nabla_{\alpha}A^{\alpha}=0.\text{\ }%
\end{equation}
The Maxwell equations read then as a wave equation for $A$, namely
\begin{equation}
\label{LorMax}
\nabla_{\alpha}\nabla^{\alpha}A^{\beta}-R^{\beta}{}_{\lambda}A^{\lambda
}=J^{\beta}=qu^{\beta},
\end{equation}
where we can replace the Ricci tensor $R_{\alpha\beta}$ by the right hand side
of (\ref{chareinstdust}).

Modulo the Maxwell equations (and $u^{\alpha}u_{\alpha}=-1),$ the stress
energy conservation equations are equivalent to
\begin{equation}
\label{charcontequ}
\nabla_{\alpha}(ru^{\alpha})=0
\end{equation}
and
\begin{equation}
\label{chargeod}
ru^{\alpha}\nabla_{\alpha}u^{\beta}+qu_{\lambda}F^{\beta\lambda}=0.
\end{equation}
Near points where $r\not =0$ equations (\ref{charcons}) and (\ref{charcontequ}%
) imply
\[
u^{\alpha}\partial_{\alpha}\left(  \frac{q}{r}\right)  =0,
\]
so that $q/r$ is constant along the flow lines. It will be constant on the
space-time to be constructed if it is constant initially. We will make this
simplifying (though not necessary) assumption, and set
\begin{equation}
\label{qrratio}
q=c\,r,
\end{equation}
with $c$ some given number. Equation (\ref{chargeod}) then implies near points
where $r>0$ holds
\begin{equation}
\label{poschargeod}
u^{\alpha}\nabla_{\alpha}u^{\beta}+cu_{\lambda}F^{\beta\lambda}=0.
\end{equation}

\subsection{Existence and uniqueness theorem.}

\begin{theorem}
The Einstein equations in wave gauge with sources an electromagnetic field of
potential in Lorentz gauge, together with a charged pressure free matter, form
a hyperbolic Leray system for $g,$ $A,$ $q,$ $u$ and $r,$ which is causal as
long as $g$ is Lorentzian and $\ u$ is timelike.\label{charexun} There is an
interval $[0,T[\,\subset\mathbb{R}$ such that the Cauchy problem for these
equations with data on the manifold $M$ such that $\bar{g}\in M_{s} $,
$\bar{A}\in H_{s}$, $\bar{k}\in H_{s-1}$, $\bar{v}\in H_{s-1}$, $\bar{r}\in
H_{s-2}$, where $s>\frac{n}{2}+2$, $|\bar{v}|_{\bar{g}}<1$ and $\bar
{q}=c\,\bar{r}$ with $c=const.$, has one and only one solution $g\in L_{s}(T)
$, $A\in H_{s}(T)$ $u\in U_{s-1}(T)$, $q=kr\in E_{s-2}(T)$.
\end{theorem}

\begin{corollary}
If the initial data satisfy the Einstein and Maxwell constraints, the solution
in wave and Lorentz gauges satisfies the original Einstein-Maxwell-dust
system. It is globally hyperbolic. There is a unique, up to isometries,
maximal, globally hyperbolic solution if $s>\frac{n}{2}+3.$
\end{corollary}

\begin{proof}
As in the previous section we differentiate the Einstein equations in wave
gauge in the direction of $u.$ We obtain, using the equation
(\ref{charcontequ}) (we could replace $u^{\gamma}\nabla_{\gamma}u_{\lambda}$
by its valued taken from (\ref{poschargeod}), but it is not necessary for our
application)
\begin{equation}
\label{derchareinstdust}
u^{\gamma}\nabla_{\gamma}R_{\alpha\beta}^{(h)}=f_{\alpha\beta},
\end{equation}
with
\begin{equation}
f_{\alpha\beta}:=u^{\gamma}\nabla_{\gamma}\Phi_{\alpha\beta}-r\{ (u_{\alpha
}u_{\beta}+\frac{1}{n - 1}g_{\alpha\beta})\,\nabla_{\gamma}u^{\gamma}
+u^{\gamma}\nabla_{\gamma}(u_{\alpha}u_{\beta})\}.
\end{equation}
We also differentiate equations (\ref{LorMax}) in the direction of $u$ to
obtain
\begin{equation}
\label{derLorMax}
u^{\gamma}\nabla_{\gamma}\nabla_{\alpha}\nabla^{\alpha}A^{\beta}-f^{\beta}%
{}_{\lambda}A^{\lambda}-R_{\lambda}^{\beta}u^{\gamma}\nabla_{\gamma}%
A_{\lambda}= c\,r\{-u^{\beta}\nabla_{\gamma}u^{\gamma}+u^{\gamma}%
\nabla_{\gamma}u^{\beta}\}.
\end{equation}

We shall consider now the system for $g$, $A$, $u$, and $r$ consisting of
equations (\ref{derchareinstdust}), (\ref{derLorMax}), (\ref{poschargeod}),
and (\ref{charcontequ}). To the equations we assign the Leray-Volevic
indices
\begin{equation}
m(\ref{derchareinstdust}) = 0, \text{ \ \ }m(\ref{derLorMax}) = 0,
\text{\ \ }m(\ref{poschargeod}) = 1, \text{ \ \ }m(\ref{charcontequ}) = 0,
\end{equation}
and to the unknowns $g$, $A$, $u$, $r$ the indices
\begin{equation}
\ell(g)=3, \text{ \ } \ell(A)=3,\text{ \ }\ell(u)=2,\text{ \ }\ell(r)=1.
\end{equation}
We see that the matrices of principal parts is again diagonal, with the same
kind of terms in the diagonal as in the previous section. The existence and
uniqueness theorem in wave and Lorentz gauge follows from the Leray-Dionne theory.

The proof of the existence of a solution of the full Einstein-Maxwell-dust
system when the initial data satisfy the Einstein and Maxwell constraints
follows the same lines as in the vacuum Einstein-Maxwell case. Geometric
global uniqueness is also proved similarly.
\end{proof}

\subsection{The motion of isolated bodies.}

Again we denote by $\omega$ the maximal open subset of $M$ on which $\bar{r} >
0$ and assume that its closure satisfies $\bar{\omega} = supp(\bar{r})$. If $\bar{r}$
vanishes on an open set so that $M \setminus\bar{\omega} \neq\emptyset$, we
denote by $\Omega$ the subset of $V = [0, T[ \times M$ which is generated by
the integral curves of $u$ which start at points of $\omega$. The support of
$r$ and $q$ is then given by the closure $\bar{\Omega}$ of $\Omega$ in $V$. It
follows that $g$ and $A$ satisfy the matter free Einstein-Maxwell equations in
$V \setminus\bar{\Omega}$.

\begin{theorem}
\label{phycharsexun}  Suppose that $\bar{r} \ge 0$. The physical solution $g$, $A$, $r,$ $q$ on $V$ and $u$ on $\Omega$ obtained in wave and Lorentz gauges in theorem 4.1 with initial
data, on $M,$ $\bar{g}$, $\bar{k}$, $\bar{A}$, $\bar{r}$, $\bar{q}=c\,\bar{r}
$ and an extension to $M$ of the datum $\bar{v}$ on ${\omega}$ with
$\bar{\omega}=supp(\bar{r})$ is independent, in a neighbourhood of $M,$ of the
extension of $\bar{v}$ to $M$.
\end{theorem}

\begin{proof}
The argument is the same as the proof of Theorem \ref{physexun} with the
geodesics curves considered there replaced by the integral curves of the
vector field $u$ defined by equation (\ref{poschargeod}).
\end{proof}

\begin{corollary}
If the initial data $\bar{g}$, $\bar{k},$ $\bar{A},$ $\bar{r},$ $\bar{q}%
=c\bar{r}$ on $M,$ \ $supp\bar{r}\subset\omega$ and $\bar{v}$ on $\bar{\omega
},$ satisfy the Einstein - Maxwell - dust constraints, then the physically
unique local solution obtained in wave and Lorentz gauge satisfies the full
Einstein- Maxwell - dust equations. It is locally geometrically unique.
\end{corollary}

\begin{proof}
The arguments are analogous to the proof of the corollary \ref{physunique}.
\end{proof}

\section{Acknowledgements}

The authors wish to thank the Isaac Newton Institute in Cambridge, where this work was
done during the programme ``Global Problems in Mathematical Relativity''.


\begin{thebibliography}{9}

%
%
%
%

\bibitem{choquet58}Choquet-Bruhat, Y. Th\'eor\`eme d'existence en m\'ecanique
des fluides relativistes. Bull. Soc. Math. France 86 (1958) 155--175.

\bibitem{dain:nagy}S. Dain, G. Nagy. Initial data for fluid bodies in general
relativity. Phys. Rev. D 65 (2002) 084020-1.

\bibitem{dionne}Dionne, P. A. Sur les probl\`emes de Cauchy hyperboliques bien
pos\'es. Journ. d'Analyses Mathematique 10 (1962) 1- 90.

\bibitem{friedrich98}Friedrich, H. Evolution equations for gravitating ideal
fluid bodies in general relativity. Phys. Rev. D 57 (1998) 2317--2322.

\bibitem{kind93}Kind, S., Ehlers, J. Initial boundary value problem for the
spherically symmetric Einstein equations for a perfect fluid. Class. Quantum
Grav. 18 (1993) 2123--2136.

\bibitem{leray53}Leray, J. (1953) Hyperbolic differential equations. Lecture
Notes, Princeton.

\bibitem{rendall:I}Rendall, A. D. The initial value problem for a class of
general relativistic fluid bodies. J. Math. Phys. 33 (1992) 1047--1053.

\bibitem{yodzis:seifert:muellerzumhagen}P. Yodzis, H.-J. Seifert, H. M\"uller
zum Hagen. On the occurrence of naked singularities in general relativity.
Commun. Math. Phys. 34 (1973) 135 - 148.
\end{thebibliography}
\end{document}